\documentclass[twocolumn,floats,showpacs,superscriptaddress,pre]{revtex4}
\usepackage{amsmath,amssymb,bm}
\usepackage{graphics}
\usepackage{epsfig}
\usepackage{graphicx}

\begin{document}

\title{Wave Functional of the Universe and Time}
\author{Natalia Gorobey}
\affiliation{Peter the Great Saint Petersburg Polytechnic University, Polytekhnicheskaya
29, 195251, St. Petersburg, Russia}
\author{Alexander Lukyanenko}
\email{alex.lukyan@mail.ru}
\affiliation{Peter the Great Saint Petersburg Polytechnic University, Polytekhnicheskaya
29, 195251, St. Petersburg, Russia}
\author{A. V. Goltsev}
\affiliation{Ioffe Physical- Technical Institute, Polytekhnicheskaya 26, 195251, St.
Petersburg, Russia}

\begin{abstract}
A version of the quantum theory of gravity based on the concept of the wave
functional of the universe is proposed. To determine the physical wave
functional, the quantum principle of least action is formulated as a secular
equation for the corresponding action operator. Its solution, the wave
functional, is an invariant of general covariant transformations of
space-time. In the new formulation, the history of the evolution of the
universe is described in terms of coordinate time together with arbitrary
lapse and shift functions, which makes this description close to the
formulation of the principle of general covariance in the classical theory
of Einstein's gravity. In the new formulation of the quantum theory, an
invariant parameter of the evolutionary time of the universe is defined,
which is a generalization of the classical geodesic time, measured by a
standard clock along time-like geodesics.
\end{abstract}

\maketitle







\section{\textbf{INTRODUCTION}}

The generally accepted and well-founded observation is considered to be the
idea of the expanding universe and the Big Bang as the beginning of this
expansion. Arising as a result of solving the equations of Einstein's
general theory of relativity (GR), it is the highest achievement of
classical physics. At the same time, the basic principles of quantum theory
were formulated as applied to atoms and molecules, and their spread to the
level of the universe as a whole was inevitable. Dirac laid the foundation
for the synthesis of relativistic and quantum principles with his exhaustive
analysis of the canonical representation of dynamical theories with singular
Lagrangian functions and their quantization \cite{Dirac}. When applied to
general relativity, this leads to the modern quantum theory of gravity (CTG)
with the Wheeler-DeWitt (WDW) system of equations \cite{Wheeler},\cite%
{DeWitt} for the wave function of the universe. This theory is based on the
canonical representation of the action of general relativity, obtained by
Arnovitt, Deser and Misner (ADM, see \cite{MTW}),
\begin{equation}
I_{ADM}=\int dt\int_{\Sigma }d^{3}x\left( \overset{\cdot }{g}_{ik}\pi ^{ik}-N%
\mathcal{H-}N_{i}\mathcal{H}^{i}\right)  \label{1}
\end{equation}%
(for the case of a closed universe), where the three-dimensional integral is
taken over the $3D$ spatial section $\Sigma $ with the metric tensor $g_{ik}$%
, $\pi ^{ik}$ is the momentum canonically conjugate to the metric, $N,N_{i}$
are arbitrary lapse and shift functions that play the role of Lagrange
multipliers, and
\begin{equation}
\mathcal{H}\left( \pi ^{ik},g_{ik}\right) =\frac{1}{\sqrt{g}}\left[ Tr%
\mathbf{\pi }^{2}-\left( Tr\mathbf{\pi }\right) ^{2}\right] -\sqrt{g}R,
\label{2}
\end{equation}
\begin{equation}
\mathcal{H}^{i}=-2\pi _{\left\vert k\right. }^{ik}  \label{3}
\end{equation}%
are Hamiltonian and momentum constraints. In this paper we use the notations
from \cite{MTW}. In this work, we do not take into account the contribution
of the matter fields. Taking it into account does not change the main
results of the work. The ADM representation is obtained using the $3+1$
-splitting of the $4D$ metric,
\begin{equation}
ds^{2}=g_{ik}\left( dx^{i}+N^{i}dt\right) \left( dx^{k}+N^{k}dt\right)
-\left( Ndt\right) ^{2}.  \label{4}
\end{equation}%
The shift functions are found by dropping the vector index of the components
$N^{k}$ of the metric: $N_{i}=g_{ik}N^{k}$. Thus, the Hamilton function in
the case of a closed universe is reduced to a linear combination of
constraints Eqs. (\ref{2}) and (\ref{3}), which in the classical theory
should be equated to zero as a consequence of the extremum of action Eq. (%
\ref{1}) with respect to $N$ and $N_{i}$. Dirac's main conclusion is that
constraints (their quantum analogs) should also vanish in quantum theory.
More precisely, we arrive at the system of WDW wave equations
\begin{equation}
\widehat{\mathcal{H}}\psi =\widehat{\mathcal{H}}^{i}\psi =0.  \label{5}
\end{equation}%
for the wave function of the universe $\psi $, which ensure its independence
from arbitrary parameters $N$ and $N_{i}$, including any external coordinate
parameter of time. The constraint operators are obtained by substituting the
canonical momenta in Eqs. (\ref{2}) and (\ref{3}) to the corresponding
differentiation operators, 
\begin{equation}
\widehat{\pi }^{ik}\left( x\right) =\frac{\hslash }{i}\frac{\delta }{\delta
g_{ik}\left( x\right) }.  \label{6}
\end{equation}%
We do not discuss here the problem of ordering non-commuting factors.

Since the wave function of the universe does not depend on any external time
parameter, the problem arises of its dynamic interpretation in terms of the
fundamental dynamic variables $g_{ik}$ (and matter fields). This problem was
solved by Hartl and Hawking using the representation of the wave function in
the form of a Euclidean functional integral \cite{HH} (no-boundary wave
function). This integral includes integration over the lapse function $N$,
and in fact over the parameter of the proper time of the universe, defined
as the integral
\begin{equation}
c=\int Ndt.  \label{7}
\end{equation}%
Thus, the wave function is also independent of the proper time Eq. (\ref{7}%
). Hartle, Hawking, and Hertog \cite{HHH,HHH1} obtained a classical picture
of an expanding universe with a scalar field within the framework of the
semiclassical approximation for a functional integral, in which the proper
time Eq. (\ref{7}) takes on the meaning of an evolution parameter. In
subsequent works \cite{HHH2}, the authors abandon the representation of the
wave function by a functional integral, formulating the semiclassical
approximation directly for the WDW equation. In \cite{GL}, a modification of
the quantization procedure and functional integral is proposed, in which the
(multi-arrow) proper time, without any approximations, acquires the meaning
of an evolution parameter. Together with its proper time, a set of
additional parameters is introduced into the theory, which form the
distribution of the own mass of the universe. However, these constructions
cannot get around the source of the time problem in modern QTG: we call the
physical state of the universe in which its Hamiltonian is zero. Thus, the
implementation of the principle of general covariance in QTG, proposed by
Dirac, leads to the emergence of the problem of time.

In this paper, a version of QTG is proposed, in which the action functional
Eq. (\ref{1}) becomes an operator in the space of wave functionals defined
on the $4D$ space-time geometries $g_{\mu \nu }$ (and matter fields). Thus,
arbitrary lapse and shift functions $N$ and $N_{i}$ are also included in
this dependence, which ensures the general covariance of the theory - the
wave functional corresponding to the physical histories of the evolution of
the universe should be an invariant of arbitrary transformations of
space-time coordinates. The quantum principle of least action (QPLA) is
formulated, according to which the quantum state of the evolution of the
universe is determined by the eigenwave functionals of the action operator,
which also depend on the boundary data on the initial and final $3D$ spatial
sections $\Sigma $. The eigenwave functional has the meaning of the
probability density for different trajectories of the evolution of the
universe in terms of the internal (coordinate time) between fixed boundary
spatial sections.

In the section $\ref{OPLA}$, QPLA will be obtained in nonrelativistic
quantum mechanics as a formulation of quantum dynamics, equivalent to the
Schr\~{o}dinger equation. 
A formulation of the QPLA in the theory of gravity is proposed. In the
section $\ref{OPLA in gravity}$, an invariant measure on the space of $4D$
geometries is introduced and the probabilistic interpretation of the wave
functional is proposed. In the section $\ref{interpretation}$, based on the
probabilistic interpretation of the wave functional, the mean proper time of
evolution of the universe between fixed boundary spatial sections is
determined. By definition, this time is an invariant of general covariant
transformations.

\section{QUANTUM PRINCIPLE OF LEAST ACTION IN NONRELATIVISTIC MECHANICS}

\label{OPLA}

To formulate new rules for quantizing the theory of gravity, let us consider
the transition to QPLA in ordinary quantum mechanics, where there is no
problem of time. It suffices to consider the simplest system with one degree
of freedom (particle) described by the Hamiltonian (we assume $m=1$)
\begin{equation}
H=\frac{p^{2}}{2}+U\left( q,t\right)  \label{8}
\end{equation}%
In ordinary quantization, canonical variables are replaced by operators (in
coordinate representation)
\begin{equation}
\widehat{q}\equiv q, \,\,\,\,\,\,\, \widehat{p}\equiv \frac{\hslash }{i}%
\frac{\partial }{\partial q},  \label{9}
\end{equation}%
which act in the space of wave functions $\psi (q,t)$. The wave function
completely defines the state of the particle at time $t$. The evolution of a
state in time is described by the Schr\"{o}dinger equation
\begin{equation}
i\hslash \frac{\partial \psi }{\partial t}=\widehat{H}\psi .  \label{10}
\end{equation}%
where
\begin{equation}
\widehat{H}=-\frac{\hslash ^{2}}{2}\frac{\partial ^{2}}{\partial q^{2}}%
+U\left( q,t\right)  \label{11}
\end{equation}%
is the Hamilton operator.

To change the quantization rules, let's look at the classical canonical form
of this theory from a different angle: we will consider the trajectory of
the particle $q(t)$ as a whole, together with the corresponding function $%
p(t)$, as a complete set of canonical variables, enumerated by a continuous
parameter $t$ \cite{JJLM}. Both of these functions are still found for given
boundary values from the condition for the extremum of the action in the
canonical form
\begin{equation}
I=\int_{0}^{T}dt\left[ \overset{\cdot }{q}p-H\left( q,p,t\right) \right] .
\label{12}
\end{equation}%
Let us formulate new quantization rules directly for action Eq. (\ref{12})
as a functional of the mentioned continuous set of canonical variables. Now,
one should expect that the state of the system, instead of the wave function
$\psi (q,t)$ given at each time instant, is described by the wave functional
$\Psi \lbrack q(t)]$, which depends on the entire trajectory. Accordingly,
we implement the generalized canonical variables by the operators
\begin{equation}
\widehat{q}\left( t\right) \equiv q\left( t\right) , \,\,\,\,\,\, \widehat{p}%
\left( t\right) \equiv \frac{\widetilde{\hslash }}{i}\frac{\delta }{\delta
q\left( t\right) },  \label{13}
\end{equation}%
acting in the space of wave functionals. Here $\widetilde{\hslash }$ is the
\textquotedblleft generalized\textquotedblright\ Planck constant, which
differs from the \textquotedblleft usual\textquotedblright\ dimension: $[%
\widetilde{\hslash }]$ = J$\cdot \sec ^{2}.$ This follows from the fact that
the variational derivative, according to its definition,
\begin{equation}
\delta \Psi =\int_{0}^{T}dt\frac{\delta \Psi }{\delta q\left( t\right) }%
\delta q\left( t\right) ,  \label{14}
\end{equation}%
has a dimension that differs from the dimension of the ordinary partial
derivative with respect to the coordinate by exactly the factor sec$^{-1}$.
We will write

\begin{equation}
\widetilde{\hslash }=\hslash \cdot \varepsilon ,  \label{15}
\end{equation}%
where $\varepsilon $ is some value of the dimension of time. The new
implementation of generalized canonical variables allows, after substituting
them in Eq. (\ref{12}), to define the action operator on the space of wave
functionals:
\begin{equation}
\widehat{I}=\int_{0}^{T}dt\left[ \frac{\widetilde{\hslash }}{i}\overset{%
\cdot }{q}\left( t\right) \frac{\delta }{\delta q\left( t\right) }+\frac{%
\widetilde{\hslash }^{2}}{2}\frac{\delta ^{2}}{\delta q^{2}\left( t\right) }%
-U\left( q\left( t\right) ,t\right) \right] .  \label{16}
\end{equation}%
Using this operator, we formulate the basic law of motion of a particle in
quantum mechanics instead of Schr\"{o}dinger equation Eq. (\ref{10}) as
follows: the physical state of motion of a particle is described by an
eigenfunctional for the action operator Eq. (\ref{16}),
\begin{equation}
\widehat{I}\Psi =\Lambda \Psi ,  \label{17}
\end{equation}%
where $\Lambda $ is the corresponding eigenvalue. This quantity, by
definition, does not depend on the specific trajectory $q(t)$, except for
its boundary points $q_{0}=q(0$) and $q_{T}=q(T)$. This statement is called
the quantum principle of least action.

Let us show that a consequence of the QPLA for the simplest theory
considered here is the Schr\"{o}dinger equation Eq. (\ref{10}). The
correspondence will be established for some singular family of wave
functionals in the limit $\varepsilon \rightarrow 0$. We define this family
as follows. Any trajectory $q(t)$ can be approximated by a broken line. For
this, we divide the time interval $[0,T]$ into small intervals of length $%
\varepsilon =T/N$ by points $t_{n}=n\varepsilon $. Take a broken line with
vertices $q_{n}=q\left( t_{n}\right) $, which is an approximation of the
trajectory $q(t)$ and coincides with it in the limit $\varepsilon
\rightarrow 0$. Let $\psi (q,t)$ be some wave function. Consider the
function of the vertices of the polyline
\begin{equation}
\Psi \left( q_{n}\right) =\prod\limits_{n=0}^{N}\psi \left(
q_{n},t_{n}\right) .  \label{18}
\end{equation}%
In the limit $N\rightarrow \infty $, we obtain the trajectory functional $%
\Psi \lbrack q(t)]$. We write the wave function in an exponential form
\begin{equation}
\psi \left( q,t\right) =\exp \left[ \frac{i}{\hslash }R\left( q,t\right) %
\right] ,  \label{19}
\end{equation}%
where $R(q,t)$ is some complex function. Then Eq. (\ref{18}) can be
represented as
\begin{equation}
\Psi \left( q_{n}\right) =\exp \left[ \frac{i}{\hslash \varepsilon }%
\sum_{n=0}^{N}\varepsilon R\left( q_{n},t_{n}\right) \right] .  \label{20}
\end{equation}%
In the limit $\varepsilon \rightarrow 0$, we obtain the wave functional of
the exponential form
\begin{equation}
\Psi \left[ q\left( t\right) \right] =\exp \left[ \frac{i}{\widetilde{%
\hslash }}\int_{0}^{T}dtR\left( q\left( t\right) ,t\right) \right] ,
\label{21}
\end{equation}%
in which, however, the infinitely small quantity $\varepsilon $ is
explicitly present. This is the singularity of the considered class of wave
functionals. Nevertheless, the operator of action Eq.(\ref{16}) on this
class of functionals will be defined insofar as it also contains this
infinitesimal quantity and their mutual cancelation will occur.

Consider the expression
\begin{equation}
\Lambda =\frac{\widehat{I}\Psi }{\Psi }  \label{22}
\end{equation}%
on an arbitrary trajectory $q(t)$ with given boundary points and find a
condition under which it does not depend on the interior points of this
trajectory. The first term in the action operator Eq. (\ref{16}) gives the
following contribution to Eq. (\ref{22}),
\begin{eqnarray}
\int_{0}^{T}dt\overset{\cdot }{q}\left( t\right) \frac{\partial R\left(
q\left( t\right) ,t\right) }{\partial q} &=&R\left( q_{T},T\right) -R\left(
q_{0},0\right)   \notag \\
&&-\int_{0}^{T}dt\frac{\partial R\left( q\left( t\right) ,t\right) }{%
\partial t}.  \label{23}
\end{eqnarray}%
The second term in Eq.(\ref{16}) requires more attention. Its calculation
rests on the expression:
\begin{equation}
\frac{1}{2}\int_{0}^{T}dt\left[ -\left( \frac{\partial R\left( q\left(
t\right) ,t\right) }{\partial q}\right) ^{2}+i\widetilde{\hslash }\frac{%
\delta \partial R\left( q\left( t\right) ,t\right) }{\delta q\left( t\right)
\partial q}\right] .  \label{24}
\end{equation}%
Here the second term will be determined if we take into account
\begin{equation}
\frac{\delta q\left( t\right) }{\delta q\left( t\right) }=\frac{\delta }{%
\delta q\left( t\right) }\int_{0}^{T}dt^{\prime }\delta \left( t-t^{\prime
}\right) q\left( t^{\prime }\right) =\delta \left( 0\right) =\frac{1}{%
\varepsilon }  \label{25}
\end{equation}%
as $\varepsilon \rightarrow 0$ (more precisely - on an interval of length $%
\varepsilon $ of our partition of the time interval $\left[ 0,T\right] $).
Then, taking into account Eq. (\ref{15}), the second term is equal
\begin{equation}
\frac{1}{2}\int_{0}^{T}dti\hslash \frac{\partial ^{2}R\left( q\left(
t\right) ,t\right) }{\partial q^{2}}  \label{26}
\end{equation}%
Putting everything together (including the contribution of the third term in
Eq. (\ref{16}), we obtain
\begin{eqnarray}
\Lambda  &=&R\left( q_{T},T\right) -R\left( q_{0},0\right)   \notag \\
&&+\int_{0}^{T}dt\left[ -\frac{\partial R\left( q\left( t\right) ,t\right) }{%
\partial t}-\frac{1}{2}\left( \frac{\partial R\left( q\left( t\right)
,t\right) }{\partial q}\right) ^{2}\right.   \notag \\
&&\left. +\frac{i\hslash }{2}\frac{\partial ^{2}R\left( q\left( t\right)
,t\right) }{\partial q^{2}}-U\left( q\left( t\right) ,t\right) \right]
\label{27}
\end{eqnarray}

It is easy to verify: if the wave function Eq. (\ref{19}) is a solution to
the Schr\"{o}dinger equation Eq. (\ref{10}), the integral in Eq. (\ref{27})
is equal to zero for any trajectory $q(t)$ and $\Lambda $ depends only on
the boundary points of this trajectory. Thus, singular exponential wave
functionals in the form Eq. (\ref{21}) satisfy the QPLA Eq. (\ref{17}) if
the wave function Eq. (\ref{19}) is a solution to the Schr\"{o}dinger
equation Eq. (\ref{10}.

The probabilistic interpretation of the wave functional $\Psi \lbrack q(t)]$
follows from the probabilistic interpretation of the wave function: it
determines the probability of a particle moving along trajectories from a
small neighborhood $q(t)$ by the formula
\begin{equation}
dP\left[ q\left( t\right) \right] =\left\vert \Psi \left[ q\left( t\right) %
\right] \right\vert ^{2}\prod\limits_{t}dq\left( t\right)  \label{28}
\end{equation}%
under the appropriate normalization condition:

\begin{equation}
\int \left\vert \Psi \left[ q\left( t\right) \right] \right\vert
^{2}\prod\limits_{t}dq\left( t\right) =1.  \label{29}
\end{equation}

\section{QUANTUM PRINCIPLE OF LEAST ACTION IN THE THEORY OF GRAVITY}

\label{OPLA in gravity}

We can transfer all the previous constructions in full to the QTG, taking as
the initial the canonical form of the ADM action Eq. (\ref{1}). As a
complete set of canonical variables, we will now consider the trajectories
of motion in time of the 3D geometry and the corresponding canonical
momentum $\left( g_{ik}\left( x,t\right) ,\pi ^{ik}\left( x,t\right) \right)
$ as a whole. Thus, time, along with spatial coordinates, now plays the role
of a numbering index. The lapse and shift functions $N$ and $N_{i}$ remain
arbitrary functions of space-time coordinates. Accordingly, we introduce
another operator realization of canonical momenta in quantum theory in the
form of partial variational derivatives with respect to time
\begin{equation}
\widehat{\pi }^{ik}\left( x,t\right) =\frac{\widetilde{\hslash }}{i}\frac{%
\delta }{\delta _{0}g_{ik}\left( x,t\right) }  \label{30}
\end{equation}%
in the space of wave functionals $\Psi \left[ g_{ik}\left( x,t\right) \right]
$. The partial variational derivative in Eq. (\ref{30}) is defined as
follows:
\begin{equation}
\delta \Psi =\int_{0}^{T}dt\int_{\Sigma }d^{3}x\frac{\delta \Psi }{\delta
_{0}g_{ik}\left( x,t\right) }\delta g_{ik}\left( x,t\right) .  \label{31}
\end{equation}%
Spatial coordinates in this definition play the role of numbering indices.
Operators of momenta of matter fields are realized in a similar way. For
comparison, we recall that in the generally accepted implementation of QTG,
the wave function of the universe $\psi $ is actually the functional of the $%
3D$ metric $g_{ik}\left( x\right) $ (and matter fields) on the spatial
section $\Sigma $, and the variational derivative is determined by the
relation:
\begin{equation}
\delta \psi =\int_{\Sigma }d^{3}x\frac{\delta \psi }{\delta _{0}g_{ik}\left(
x\right) }\delta g_{ik}\left( x\right) .  \label{32}
\end{equation}%
In such an implementation, modification Eq. (\ref{15}) of the Planck
constant is not required. Substituting the operators of momenta Eq. (\ref{30}%
) into action Eq. (\ref{1}), we obtain the operator of the action of the ADM
in the modified QTG:
\begin{equation}
\widehat{I}_{ADM}=\int dt\int_{\Sigma }d^{3}x\left( \overset{\cdot }{g}_{ik}%
\widehat{\pi }^{ik}-N\widehat{\mathcal{H}}\mathcal{-}N_{i}\widehat{\mathcal{H%
}}^{i}\right) ,  \label{33}
\end{equation}%
where, as we agreed, the momentum operators Eq. (\ref{30}) in all
expressions are on the right. Now let us formulate the quantum principle of
least action in the theory of gravity as a spectral problem for the action
operator:
\begin{equation}
\widehat{I}_{ADM}\Psi =\Lambda _{ADM}\Psi .  \label{34}
\end{equation}%
Here the eigenvalue $\Lambda _{ADM}$ is a (complex) function of $3D$
geometries $g_{ik}^{0},g_{ik}^{T}$ on two boundary spatial sections $\Sigma
_{0},\Sigma _{T}$ of $4D$ Lorentz space-time. The eigenvector - the wave
functional $\Psi $ depends on all components of the $4D$ space-time metric
between the boundary spatial sections, including arbitrary Lagrange
multipliers $N$ and $N_{i}$. The desired wave functional $\Psi \left[ g_{\mu
\nu }\left( x,t\right) \right] $, which describes the physical history of
the evolution of the universe, must be an invariant of general covariant
transformations and this is ensured by its dependence on $N$ and $N_{i}.$

\section{INTERPRETATION OF THE WAVE FUNCTIONAL OF THE UNIVERSE AND TIME}

\label{interpretation}

As it should be in quantum theory, the wave functional describes the
evolution of the universe in a probabilistic way. But this description must
be covariant. Since the wave functional itself is an invariant, one should
also introduce an invariant measure of integration on the set of $4D$
space-time geometries between the given boundary spatial sections $\Sigma
_{0},\Sigma _{T}$. Such a measure was constructed in the works of Faddeev
and Popov \cite{FP,KP} and here we will use the ready result:
\begin{eqnarray}
d\mu _{FP}\left[ g_{\mu \nu }\right] &=&\Delta _{h}\left[ g_{\mu \nu }\right]
\times  \notag \\
&&\times \prod\limits_{x}\left( \prod\limits_{\beta }\delta \left( \partial
_{\alpha }\left( \sqrt{-g}g^{\alpha \beta }\right) \right) \times \right.
\notag \\
&&\left. \times \left( -g\right) ^{\frac{5}{2}}\prod\limits_{\mu \leqslant
\nu }dg^{\mu \nu }\right) ,  \label{35}
\end{eqnarray}%
where additional gauge harmonicity conditions are used, and $\Delta _{h}%
\left[ g_{\mu \nu }\right] $ is the corresponding Faddeev-Popov determinant.
Thus, the probability that the universe evolved in a small neighborhood of a
given $4D$ space-time geometry $g_{\mu \nu }(x,t)$ is given by the
expression
\begin{equation}
dP\left[ g_{\mu \nu }(x,t)\right] =\left\vert \Psi \left[ g_{\mu \nu }\left(
x,t\right) \right] \right\vert ^{2}d\mu _{FP}\left[ g_{\mu \nu }\left(
x,t\right) \right]  \label{36}
\end{equation}%
with the additional normalization condition:
\begin{equation}
\int \left\vert \Psi \left[ g_{\mu \nu }\left( x,t\right) \right]
\right\vert ^{2}d\mu _{FP}\left[ g_{\mu \nu }\left( x,t\right) \right] =1.
\label{37}
\end{equation}
Now we can answer the question: what physical (invariant) time separates the
boundary spatial sections $\Sigma _{0},\Sigma _{T}$? Here the parameter $T$
is an arbitrary coordinate time, which is fixed by us for convenience. As a
physical measure of time between two spatial sections, it is natural to take
geodesic time, measured by a clock moving along a time-like geodesic. More
precisely, in a given $4D$ space-time geometry $g_{\mu \nu }\left(
x,t\right) $, we take a family of geodesics orthogonal to the initial
spatial section $\Sigma _{0}$ and calculate the average over this family
(integral over $\Sigma _{0}$) the geodesic distance between the boundary
spatial sections $\Sigma _{0},\Sigma _{T}$:
\begin{equation}
\left\langle \int_{\Sigma _{0}}^{\Sigma _{T}}\sqrt{-g_{\mu \nu }dx^{\mu
}dx^{\nu }}\right\rangle _{\Sigma _{0}}.  \label{38}
\end{equation}%
Now this value should be averaged over all possible $4D$ space-time
geometries between the given boundary spatial sections $\Sigma _{0},\Sigma
_{T}$:
\begin{equation}
S=\int \left\langle \int_{\Sigma _{0}}^{\Sigma _{T}}\sqrt{-g_{\mu \nu
}dx^{\mu }dx^{\nu }}\right\rangle _{\Sigma _{0}}d\mu _{FP}\left[ g_{\mu \nu }%
\right] .  \label{39}
\end{equation}%
Integral Eq. (\ref{39}) gives the mathematical expectation for the proper
time between two boundary states of the universe on the spatial sections $%
\Sigma _{0},\Sigma _{T}$.

The construction proposed here also allows one to determine an analogue of
the no-boundary of the Hartle and Hawking wave function. To do this, one
should go to the Euclidean form of action of ADM, Eq. (\ref{1}), and QPLA,
Eq. (\ref{34}). The transition to the Euclidean $4D$ metric in
representation Eq. (\ref{4}) is achieved by a simple replacement
\begin{equation}
N\longrightarrow iN.  \label{40}
\end{equation}%
The canonical momentum of the $3D$ ADM metric, according to its definition
(see \cite{MTW}), changes as follows: $\mathbf{\pi }\rightarrow -i\mathbf{%
\pi }$. As a result, the canonical action of ADM is transformed into the
Euclidean form: $I_{ADM}\longrightarrow -iI_{ADM}^{E}$. In the Euclidean
form of QPLA, one can search for a solution in the form of a real wave
functional defined on $4D$ Euclidean geometries with one boundary spatial
section or no boundaries at all. The return to the Lorentzian signature of
the $4D$ metric is carried out in the found solution by rotating in the $N$
complex plane (where possible), inverse to Eq. (\ref{40}).

In conclusion, we note that the singular character of the QPLA in this
formulation, which assumes the passage to the limit $\varepsilon \rightarrow
0$ in Eq. (\ref{15}), admits regularization by adding to the
Hilbert-Einstein action with the Lagrange function $-1/\varkappa ^{2}\sqrt{-g%
}R$ the square of scalar curvature (the theory with the Lagrangian function $%
-1/\varkappa ^{2}R+\sigma ^{2}R^{2}$ ). In this theory, the Lagrange
function contains second-order derivatives of the metric. Its canonical
analysis and standard quantization are carried out in \cite{MN} based on the
approach developed by Ostrogradskii for theories with higher derivatives
\cite{O}. The modified Lagrange function contains an additional dimensional
parameter $\varkappa \sigma $, which serves as a measure of the theory's
nonlocality. Within the framework of the approach proposed in this paper to
the formulation of the QPLA, the nonlocality of the dynamics does not
prevent the possibility of determining the set of generalized canonical
variables formed by their values at all times. In this case, the canonical
momenta conjugate to the metric $g_{ik}\left( x,t\right) $ should be defined
as the variational derivatives of the action with respect to the velocities $%
\overset{\cdot }{g}_{ik}\left( x,t\right) $. The resulting canonical action
will be nonlocal, and in the QPLA formulation, the modified Planck constant
should be set equal to
\begin{equation}
\widetilde{\hslash }=\hslash \cdot \varkappa \sigma .  \label{41}
\end{equation}%
In the limit $\sigma \rightarrow 0$, we return to the original singular form
of the QPLA.

\section{CONCLUSIONS}

The wave function in the generally accepted QTG, if we follow the canonical
structure of the ADM action Eq. (\ref{1}), describes the state of the
universe at a particular moment of coordinate time. However, the formulation
of the principle of general covariance in quantum theory, according to
Dirac, requires the removal of this time parameter from consideration as
non-physical. And this is achieved by the conditions for the vanishing of
quantum constraints. In this case, the wave function contains information
about the history of the evolution of the universe only in integral form in
its phase. Recovery of this information is possible in the semiclassical
approximation, and this procedure is called the holographic principle in
\cite{HHH1,HHH2}. The introduction of the wave functional, which is an atlas
of the history of the universe in coordinate time, allows us to speak about
this history explicitly. QPLA allows one to define a solution that is
invariant with respect to general covariant transformations. Thus, an
alternative embodiment of the principle of general covariance in the quantum
theory of gravity is proposed, which includes coordinate time together with
arbitrary lapse and shift functions. This implementation is close to its
formulation in the classical GR of Einstein. This makes it possible to
determine in the quantum theory the invariant parameter of the evolutionary
time of the universe, which generalizes the concept of proper time measured
along time-like geodesics.

\section{ACKNOWLEDGEMENTS}

We are thanks V.A. Franke for useful discussions.




\end{document}